\documentclass[showpacs,preprintnumbers,amsmath,amssymb,superscriptaddress]{revtex4}
\usepackage{amssymb}
\usepackage{mathrsfs}
\usepackage{graphicx}
\usepackage{dcolumn}
\usepackage{bm}
\usepackage{mathrsfs}

\begin{document}

\title{Astrophysical uncertainties on the local dark matter distribution and direct detection experiments}

\author{Anne M. Green}\affiliation{School of Physics and Astronomy, University of Nottingham,
  University Park, Nottingham, NG7 2RD, UK\\
anne.green@nottingham.ac.uk}


\begin{abstract}
 The differential event rate in Weakly Interacting Massive Particle (WIMP) direct detection experiments depends on the local dark matter density and velocity distribution. Accurate modelling of the local dark matter distribution is therefore required to obtain reliable constraints on the WIMP particle physics properties. Data analyses typically use a simple Standard Halo Model which might not be a good approximation to the real Milky Way (MW) halo. We review observational determinations of the local dark matter density, circular speed and escape speed and also studies of the local dark matter distribution in simulated MW-like galaxies. We discuss the effects of the uncertainties in these quantities on the energy spectrum and its time and direction dependence. Finally we conclude with an overview of various methods for handling these astrophysical uncertainties.  
\end{abstract}

\date{\today}
\maketitle


\section{Introduction}	

Direct detection experiments aim to detect cold dark matter in the form of Weakly Interacting Massive Particles (WIMPs) via their elastic scattering off target nuclei~\cite{Goodman:1984dc}. The signals expected in these experiments, and hence the resulting constraints on the WIMP mass and cross-sections, depend on the local WIMP density and velocity distribution. We first outline the expressions for the differential event rate, including its time and direction dependence, focussing on how it depends on these astrophysical quantities (Sec.~\ref{background}). In Secs.~\ref{astro1} and \ref{astro2} we review what observations and simulations tell us about the local dark matter distribution, before discussing the effect of astrophysical uncertainties on the signals expected in direct detection experiments in Sec.~\ref{cons}. Finally, in Sec.~\ref{strategies}, we discuss strategies for handling astrophysical uncertainties, including `astrophysics independent' methods, parameterizing the WIMP speed distribution and building self-consistent models of the Milky Way (MW).

\section{Background}
\label{background}

\subsection{Differential event rate}
The differential event rate, per unit detector mass, as a function of recoil energy, $E$, and direction, $\Omega$, under the standard assumption of elastic scattering due to contact interactions, is given by
\begin{equation}
\label{drdedir}
\frac{{\rm d}^2 R}{{\rm d} E \, {\rm d} \Omega} = \frac{ \rho_{0}}{4 \pi m_{\chi} \mu_{\chi A}^2} \left[ \sigma^{\rm SI} F_{\rm SI}^2 (E) + \sigma^{\rm SD} F_{\rm SD}^2 (E) \right] \hat{f}_{\rm lab}(v_{\rm min}, \hat{{\bf r}}) \,,
\end{equation}
where $\rho_{0}$ is the local (i.e. at the Solar radius) dark matter density, $m_{\chi}$ is the WIMP mass, $\mu_{\chi A}$ the WIMP-nucleus reduced mass, $\sigma^{\rm SI , \, SD}$ the spin independent/spin dependent WIMP-nucleus cross section at zero momentum transfer, $F_{\rm SI, \, SD}(E)$ the   spin independent/spin dependent form factor, $\hat{{\bf {r}}}$ a unit vector in the direction of the recoiling nucleus 
\begin{equation}
v_{\rm min} = \left( \frac{ E m_{A}}{2 \mu_{\chi A}^2} \right)^{1/2} \,,
\end{equation}
is the minimum WIMP speed that can cause a recoil of energy $E$ and $m_{A}$ is the mass of the nucleus. Finally $\hat{f}_{\rm lab}(v_{\rm min}, \hat{{\bf r}})$ is the Radon transform of the lab frame WIMP velocity distribution, $f_{\rm lab}({\bf v})$,
\begin{equation}
\hat{f}_{\rm lab}(v_{\rm min}, \hat{{\bf r}}) = \int \delta ({\bf v}. \hat{{\bf r}} - v_{\rm min} ) f_{\rm lab}({\bf v}) \, {\rm d}^3 {\bf v} \,, 
\end{equation}
i.e.~the integral of $f_{\rm lab}({\bf v})$ over the plane orthogonal to the direction $\hat{{\bf r}}$ a distance $v_{\rm min}$ from the origin.
For the majority of direct detection experiments, which can measure the energy of the nuclear recoils but not their directions, Eq.~(\ref{drdedir}) becomes
\begin{equation}
\label{drde}
\frac{{\rm d} R}{{\rm d} E } = \frac{ \rho_{0}}{2 m_{\chi} \mu_{\chi A}^2} \left[ \sigma^{\rm SI} F_{\rm SI}^2 (E) + \sigma^{\rm SD} F_{\rm SD}^2 (E) \right] 
g(v_{\rm min})  \,,
\end{equation}
where $g(v_{\rm min})$ is the velocity integral:
\begin{equation}
\label{gvmin}
g(v_{\rm min})  = \int_{v> v_{\rm min}}^{\infty} \frac{ f_{\rm lab}({\bf v})}{v} \, {\rm d}^3 {\bf v} \,.
\end{equation}
Since $f({\bf v})$ is positive definite, $g(v_{\rm min})$ must be a monotonically decreasing function of $v_{\rm min}$. The spin independent cross section can be written as
\begin{equation}
\sigma^{\rm SI} = \frac{4}{\pi} \mu_{\chi A}^2 \left[ Z f_{p} + (A-Z) f_{n} \right]^2 \,,
\end{equation}
where $Z$ and $A$ are the atomic and mass number of the nucleus and $f_{p,n}$ is the coupling between the WIMP and protons/neutrons. The cross section for scattering on a proton is
\begin{equation}
\sigma_{p}^{\rm SI} = \frac{4}{\pi} \mu_{\chi p}^2 f_{p}^2 \,,
\end{equation}
where $\mu_{\chi p}$ is the WIMP-proton reduced mass and in the case where  $f_{p} = f_{n}$ the total spin independent cross section can be written as
\begin{equation}
\sigma^{\rm SI} = \left( \frac{\mu_{\chi A}}{\mu_{\chi p}} \right)^2 A^2 \sigma_{p}^{\rm SI} \,.
\end{equation}
For further details of the derivation of Eqs.~(\ref{drdedir}) and (\ref{drde}) see e.g. Ref.~\cite{Gondolo:2002np}.

The lab frame speed distribution $f_{\rm lab}(\bf v)$ is time dependent, due to the time dependent transformation from the Galactic rest frame to the lab frame:  ${\bf v} \rightarrow  {\bf v} + {\bf v}_{e}(t)$. The Earth's motion relative to the Galactic rest frame, ${\bf v}_{e}(t)$, is made up of three
components: the motion of the Local Standard of Rest (LSR)~\footnote{This assumes the the Local Standard of Rest is identical to the Rotational Standard of Rest, this might not be the case~\cite{Bovy:2012ba}.}, ${\bf  v}_{\rm LSR}=(0,v_{\rm c},0)$~\footnote{The coordinate system is defined so that the first axis points towards the Galactic Center, the second along the direction of Galactic rotation and the third is perpendicular to the Galactic plane.} where $v_{\rm c}$ is the local circular speed (see Sec.~\ref{vc}), the Sun's peculiar motion with respect to the LSR, ${\bf v}_{\odot}^{\rm p}$, (see Sec.~\ref{vp}) and the Earth's orbit about the Sun, ${\bf v}_{e}^{\rm orb}$~\footnote{For an correct, accurate expression for Earth's orbital speed
see Refs.~\cite{McCabe:2013kea,Lee:2013xxa}}. The Earth's orbit leads to an annual modulation in the differential event rate~\cite{dfs}. The recoil rate is also strongly direction dependent. The recoils are tightly concentrated around the inverse of the direction of Solar motion, such that the recoil rate in the forward hemisphere is roughly an order of magnitude larger than that in the backwards one~\cite{Spergel:1987kx}.

The interaction between the WIMP and the nucleus may also be inelastic (due to, for instance, the WIMP particle belonging to a multiplet of states)~\cite{TuckerSmith:2001hy}. In this case the relationship between $v_{\rm min}$ and $E$, and hence the dependence of the differential event rate on the velocity distribution, is modified. There has also recently been interest in using non-relativistic effective operators to consider a wider range of WIMP-nucleus interactions, and again the relationship between the velocity distribution and the event rate is changed~\cite{Fan:2010gt,Fitzpatrick:2012ix}.

\subsection{Phase space distribution function} 
\label{psdf}

The steady-state phase space distribution, $f({\bf x}, {\bf v})$, of a collection of collisionless particles is given by the solution of the collisionless Boltzmann equation: ${\rm d} f/ {\rm d} t=0$ (see e.g.~Ref.~\cite{bt}). The standard halo model (SHM) is an isotropic, isothermal sphere with density profile $\rho(r) \propto r^{-2}$. In this case the solution to the collisionless Boltzmann equation is a Maxwellian velocity distribution
\begin{equation}
f({\bf v}) = \frac{N}{2 \pi \sigma_{v}^2} \exp{ \left( - \frac{{\bf v}^2}{2 \sigma_{v}^2} \right) } \,,
\end{equation}
where $N$ is a normalization constant and $\sigma_{v}$ is the one-dimensional velocity dispersion, which in the SHM is related to the circular speed, $v_{\rm c}$, (see Sec.~\ref{vc}) by $\sigma_{v} = v_{\rm c} / \sqrt{2}$. The circular speed  depends on the potential, $\Phi(r)$:  $v_{\rm c}^2= r {\rm d} \Phi / {\rm d} r$ (see e.g.~Ref.~\cite{bt}).
Therefore, as emphasised in Ref.~\cite{Peter:2011eu}, the relationship between the one-dimensional velocity dispersion and the circular speed depends on the density profile. For a power law density profile, $\rho(r) \propto r^{-\beta}$, $\sigma_{v} = v_{\rm c} / \sqrt{\beta}$. Formally the Maxwellian velocity distribution extends to infinity, therefore it is usually truncated by hand at the escape speed, $v_{\rm esc}$ (Sec.~\ref{vesc}). 
 
For spherical isotropic systems the Eddington equation (see e.g.~Ref.~\cite{bt}) provides a unique relationship between the density profile and the distribution function. However for triaxial and anisotropic systems there is no unique relationship between the density profile and the velocity distribution.  
For an anisotropic system the phase space distribution function is a function of both energy and the magnitude of the angular momentum. In this case solutions can be found using specific ansatzes for the form of the anisotropy~\cite{o,m,Bozorgnia:2013pua,Fornasa:2013iaa}. Another approach, used in Ref.~\cite{Evans:2000gr} for triaxial systems, is to use the Jeans equations (which are derived by taking moments of the collisionless Boltzmann equation) to find the components of the velocity dispersion and approximate the velocity distribution as a multi-variate gaussian.

The formation of dark matter halos is hierarchical, rather than monolithic, and therefore additional non-virialised contributions to the phase space distribution function, such as tidal streams, are possible. We discuss these features in Sec.~\ref{astro2}.

\section{Observations}
\label{astro1}

\subsection{Density}

In the context of WIMP direct detection experiments, the standard value of the local dark matter density is $\rho_{0} = 0.3 \, {\rm GeV \, cm}^{-3}$ (or equivalently $0.008 M_{\odot} \, {\rm pc}^{-3}$). Observational probes of the dark matter density at the Solar radius can be divided into two classes: local and global. Local measures use the vertical motion of tracer stars in the Solar neighbourhood, while global measures use an ensemble of data sets to constrain a parameterized model of the MW.  The local dark matter density could in principle be larger than the global (spherically averaged) value if the MW halo is flattened or there is a dark disc. For a detailed review see Ref.~\cite{Read:2014qva}.


There is a long history of local measurements of the local dark matter density, dating back to Kapteyn~\cite{kapteyn}, Jeans~\cite{jeans} and Oort~\cite{oort} in the 1920s and 30s.
In recent years there have been a number of studies, using different data sets and different methodologies (and hence different assumptions). These studies agree with each other within their quoted uncertainties, finding best-fit values in the range $(0.22-0.33) \, {\rm GeV \, cm}^{-3}$~\cite{bovyt,garbari1,gabari2,zhang,bovyr,Silverwood:2015hxa}. The errors depend on the method used and lie, roughly, in the range ($0.05-0.5)  \, {\rm GeV \, cm}^{-3}$. See Fig. 2, Table 4 and Sec. 5.2 of Ref.~\cite{Read:2014qva} for further details. Recent determinations of the global dark matter density typically lie in the range ($0.2-0.6)  \, {\rm GeV \, cm}^{-3}$~\cite{salucci,catenaullio,weberdeboer,Iocco,mcmillan,Pato:2015dua,huang,mcmillan2}. The size of the statistical errors (which in some cases are as small as $10\%$) depend on the assumptions made.
See Table 4 and Sec. 5.3 of Ref.~\cite{Read:2014qva} for further details.

As discussed in Ref.~\cite{Read:2014qva},  significant improvement in the accuracy (and precision) of determinations of the local dark matter density is expected in the near future using data from the Gaia satellite.

\subsection{Circular speed}
\label{vc}

The standard value of the local circular speed is $v_{\rm c} = 220 \, {\rm km \, s}^{-1}$. As discussed in Sec.~5.2 of Ref.~\cite{Bovy:2012ba}, the local circular speed can be measured via two different general methods: by measuring the Sun's velocity with respect to an object assumed to be at rest with respect to the Galactic centre or by directly measuring the local radial force (via the Oort constants or observations of tidal streams). 

The measured proper motion of Sgr $A^{\star}$~\cite{Reid:2004rd}, combined with a Solar radius of $R_{0} \approx 8 \, {\rm kpc}$, gives a total Solar velocity of $v_{{\phi}, \odot} = 242 \, {\rm km \, s}^{-1}$. The $\phi$ component of the Sun's motion with respect to the Rotational Standard of Rest (RSR) then has to be subtracted to find $v_{\rm c}$.  Ref.~\cite{Bovy:2012ba} measures the Milky Way rotation curve using stellar tracers. They find $v_{\rm c} =(218 \pm 6) \, {\rm km \, s}^{-1}$ and $v_{{\phi}, \odot} = 242_{-3}^{+10} \, {\rm km \, s}^{-1}$, with a value for the $\phi$ component of the Sun's velocity relative to the RSR of $\sim 24 \, {\rm km \, s}^{-1}$ which is somewhat larger than the $\phi$ component of the Sun's velocity with respect to the LSR~\cite{Schoenrich:2009bx}. Ref.~\cite{Koposov:2009hn} finds $v_{\rm c} = 220 \pm 18 \, {\rm km \, s}^{-1}$ from the orbit of the GD-1 stellar stream. A larger value of $v_{\rm c} = (254 \pm 16) \, {\rm km \, s}^{-1}$ was claimed using the kinematics of masers in the Galactic disk~\cite{Reid:2009nj}, however subsequent reanalyses, taking into account the non-random orbital phases of the masers, have found a lower central value and significantly larger uncertainties~\cite{Bovy:2009dr,McMillan:2009yr}. The local circular speed is proportional to the gradient of the potential, and hence depends on the density profile (see Sec.~\ref{psdf}).

\subsection{Peculiar motion of the Sun}
\label{vp}
The components of Sun's peculiar motion with respect to the LSR have been determined to be ${\bf v}_{\odot}^{\rm p} = (11.1_{-0.8}^{+0.7},\, 12.2_{-0.5}^{+0.5},\, 7.3_{-0.4}^{+0.4}) \, {\rm km \, s}^{-1}$ with systematic uncertainties of order $(1,\,2,\,0.5) \, {\rm km \, s}^{-1}$~\cite{Schoenrich:2009bx}.  

\subsection{Escape speed}
\label{vesc}

The escape speed is the speed above which stars and particles are not gravitationally bound to the MW: $v_{\rm esc}(r)= \sqrt{ 2 |\Phi(r)|}$ where $\Phi(r)$ is the potential. The local escape speed, $v_{\rm esc} \equiv v_{\rm esc}(r=R_{0})$, is estimated from the speeds of local high velocity halo stars. 

The most recent estimate, $v_{\rm esc}= 533_{-41}^{+54} \, {\rm km \, s}^{-1}$ at 90\% confidence, comes from Ref.~\cite{Piffl:2013mla} using data from the Radial Velocity Experiment (RAVE). They define the escape speed as the minimum speed required to reach three virial radii~\footnote{Ref.~\cite{Piffl:2013mla} takes the virial radius to be the radius which encloses a mean density which is $340$ times the critical density for which the geometry is flat, where $340$ is the calculated virial over-density for a flat universe with 30\% of the total density in matter (and the remainder in a cosmological constant).} and parametrise the tail of the stellar velocity distribution as $f(v) \propto (v_{\rm esc} - v)^{k}$~\cite{lt}, which Ref.~\cite{Piffl:2013mla} found to be a good fit to cosmological simulations with $k$ in the range $2.3< k< 3.7$. Note that, as emphasised by Ref.~\cite{Lavalle:2014rsa}, the escape speed is correlated with other astrophysical parameters, in particular the circular speed, $v_{\rm c}$.

\section{Simulations}
\label{astro2}
Since the local dark matter distribution can not be directly `observed' (apart from by lab based dark matter detection experiments!) numerical simulations are a useful source of information about this quantity. For a detailed review of the state of the art (as of 2012) and future prospects see Ref.~\cite{Kuhlen:2012ft}.

The resolution of simulations is many orders of magnitude larger than the scales probed by Earth based experiments (the Earth moves at $\sim 200 \, {\rm km} \, {\rm s}^{-1} \sim 0.1 \, {\rm mpc} \, {\rm yr}^{-1}$ with respect to the Galactic rest frame). This raises the question of whether the dark matter distribution on these scales contains ultra-fine structure that can not be resolved by simulations. Vogelsberger and White~\cite{Vogelsberger:2010gd} addressed this issue by devising a technique which probes the evolution of the fine-grained phase space distribution within a numerical simulation. They found that the typical density of the resulting streams is of order $10^{-7}$ times the local halo density, and hence the local dark matter distribution consists of a huge number of overlapping streams, producing a close to smooth distribution. Similar conclusions have been reached by Schneider et al.~\cite{Schneider:2010jr} by considering the evolution of the first WIMP microhalos~\cite{Hofmann:2001bi,Green:2003un,Green:2005fa,Diemand:2005vz} that form.

\subsection{N-body}

In the 2000s various high resolution, dark matter-only simulations of MW-like halos in a cosmological context were carried out (e.g. Aquarius~\cite{aquarius}, GHALO~\cite{ghalo} and Via Lactea~\cite{vl}). The local velocity distributions of these halos were found to deviate systematically from a multi-variate Gaussian, having more low speed particles and a lower, flatter peak~\cite{hansen,fairs,vogelsberger,kuhlen}.  These distributions are well fit by a phenomenological form proposed by Mao et al.~\cite{Mao:2012hf}:
\begin{equation}
\label{maofv}
f(|{\bf v}|) = 
\begin{cases}
 N 
 \exp{\left( - \frac{|{\bf v}|}{v_{0}} \right)}  \left( v_{\rm esc}^2 - |{\bf v}|^2 \right)^p   &
0 \leq |{\bf v}|  \leq v_{\rm esc} \,, \\
 0 \,, &  {\rm otherwise} \,.
\end{cases}
\end{equation}
The parameter ($v_{0}/v_{\rm esc}$) varies with radius in a single halo and $p$ lies in the range $[0,3]$ with significant scatter from halo-to-halo~\cite{Mao:2012hf,Mao:2013nda}.

In additional to the dominant smooth component, the dark matter distribution contains stochastic components, in particular at high speed. At some locations there are tidal streams, giving narrow spikes in the velocity distribution~\cite{kuhlen}. There are also broad features, which vary from halo to halo, but do not vary with position within a single halo. These features, which have been dubbed `debris flow', are believed to reflect the formation history of the halo~\cite{Lisanti:2011as,Kuhlen:2012fz}.

\subsection{Hydrodynamical}

The contribution of baryons and dark matter to the potential in the Solar neighbourhood is comparable. Therefore the baryons are likely to have a non-negligible effect on the local dark matter distribution. Simulating baryonic physics requires formulating prescriptions for physical processes which occur on scales that are smaller than the resolution of the simulation. This is extremely challenging, however there have been significant advances in recent years in producing simulated galaxies which resemble the Milky Way.

Refs.~\cite{ling1,Kuhlen:2013tra,Butsky:2015pya} found velocity distributions in hydrodynamical simulations of MW-like galaxies which deviated significantly, in different ways, from a Maxwellian distribution. In early 2016 several studies of the local dark matter distribution in MW-like galaxies were published~\cite{Bozorgnia:2016ogo,Kelso:2016qqj,Sloane:2016kyi}. Borzognia et al.~\cite{Bozorgnia:2016ogo} selected galaxies from the EAGLE~\cite{Crain:2015poa} and APOSTLE~\cite{apostle} projects with masses in the range ($10^{12}-10^{13} \, M_{\odot}$) that satisfied observational constraints on the MW rotation curve and stellar mass. 
Kelso et al.~\cite{Kelso:2016qqj} studied two galaxies with global properties similar to the MW from the MaGICC~\cite{Stinson:2012uh} simulations.
Sloane et al.~\cite{Sloane:2016kyi} studied four MW-like galaxies, with masses in the range $0.7-0.9 \times 10^{12} \, M_{\odot}$ with a variety of different merger histories (i.e. quiescent or with one or more relatively recent major mergers). The resolution of their simulations is higher than those studied in Ref.~\cite{Bozorgnia:2016ogo,Kelso:2016qqj}.

In general the baryons deepen the potential and increase the average speed of the dark matter particles. Refs.~\cite{Bozorgnia:2016ogo,Kelso:2016qqj} found that a Maxwellian speed distribution is a good fit to the speed distribution in their hydrodynamical simulations. This may be because adiabatic contraction makes the logarithmic slope of the density profile in the Solar neighbourhood closer to that of an isothermal sphere~\cite{Kelso:2016qqj}. 
Even though the Maxwellian speed distribution is a good fit, there is still significant halo to halo variation in the peak speed/velocity dispersion. Ref.~\cite{Sloane:2016kyi} also found that a Maxwellian speed distribution is a better fit to their hydrodynamical simulations than to their dark matter only simulations, however they still found a deficit of high speed particles relative to the Maxwellian. They also found that the Earth frame speed distribution in the hydrodynamical simulations contained features and the velocity distributions varies more between halos in the hydrodynamical simulations than in the dark matter only ones.

In some baryonic simulations a co-rotating dark matter disc is formed, due to late merging sub-halos being preferentially dragged towards the stellar disc by dynamical friction~\cite{read1,read2,ling1}. Ref.~\cite{purcell} argued that to be consistent with the observed properties of the MW's thick disc, the MW's merger history must be quiescent compared with typical $\Lambda$CDM merger histories, and hence the density of its dark disc must be relatively small. Most of the MW-like galaxies formed in recent hydrodynamical simulations do not have a significant dark disc~\cite{Bozorgnia:2016ogo,Kelso:2016qqj}.

Ref.~\cite{Purcell:2012sh} studied the ongoing disruption of the Sagittarius dwarf galaxy, and in particular the question of whether one of its tidal streams passes through the Solar neighbourhood (as suggested, in the context of direct detection, by Ref.~\cite{Freese:2003tt}). They found that the dark matter in the stream is more extended, and not coaxial with, the stellar stream and can hence have a non-negligible contribution to the local dark matter distribution even if the stellar stream does not pass through the Solar neighbourhood. Ref.~\cite{Petersen:2016vck} studied the dynamical effects of a galactic bar on the dark matter distribution, finding that the DM density at the Solar radius varies depending on the Earth's location relative to the stellar bar and a quadrupole wake in the DM that lags the stellar bar forms.

In summary the simulations discussed above use different prescriptions for sub-grid physics as well as different criteria for selecting MW-like galaxies. Recent studies~\cite{Bozorgnia:2016ogo,Kelso:2016qqj,Sloane:2016kyi} suggest that the speed distribution does not deviate hugely from a Maxwellian, however there may still be non-negligible deviations and a firm consensus about the exact form of the local dark matter velocity distribution has still to emerge. In particular there may be significant variation from halo to halo.

\section{Consequences}
\label{cons}

The normalization of the differential event rate is proportional to the product of the density and cross-section, so the accuracy of constraints on the cross-section is limited by the accuracy with which the local density is known. The differential event rate is proportional to an integral over the velocity distribution, Eq.~(\ref{drde}), and hence it depends fairly weakly on the detailed shape of the velocity distribution~\cite{kk,donato}. Therefore the resulting uncertainty in exclusion limits~\cite{greenexclude,mccabe,Fairbairn:2012zs,Benito:2016kyp} and determinations of the WIMP mass~\cite{wimpmassme1,wimpmassme2} is usually fairly small. The main exception to this is the case of light WIMPs.  If the WIMP is much lighter than the target nuclei, or the experimental energy threshold is large, then the experiment is only sensitive to the high speed tail of the speed distribution. In this case the precise value of the escape speed, and also the shape of the speed distribution just below it, can have a significant effect on the energy spectrum, and hence exclusion limits or determinations of the WIMP parameters~\cite{mccabe,lsww}. Astrophysical uncertainties also affect the level of the `neutrino floor'~\cite{OHare:2016pjy}, the cross-section below which neutrino backgrounds strongly limit the sensitivity of non-directional experiments~\cite{Billard:2013qya}.

If the WIMP velocity distribution is isotropic then the phase of the annual modulation (defined as the time at which the event rate is largest for large energies), $t_{0}$, is determined by when the Earth's velocity with respect to the Galactic rest frame is largest. In this case $t_{0} \approx 150$ days, with the exact value depending on the Sun's peculiar velocity~\cite{greenam2}. The annual modulation arises from the small shift in the speed distribution in the lab frame due to the Earth's orbit. The properties of the annual modulation are therefore more sensitive to the speed distribution than the time averaged differential event rate. The amplitude of the modulation, and hence the regions of parameter space compatible with an observed signal, can vary significantly~\cite{br,vergados1a,vergados1b,belli1,greenam1,vergados2,greenam2,belli2,vergados3,fs,kuhlen,greenfv,cbc,Kuhlen:2013tra}. 
For analytic anisotropic velocity distributions the phase of the modulation can shift by up to tens of days~\cite{fs,Copi:2002hm,greenfv}. Some simulations have found an energy dependent shift in the phase of up to five days~\cite{kuhlen,Kuhlen:2013tra}. The detailed angular dependence of the directional event rate is sensitive to the exact shape of the velocity distribution, however the `forwards-backwards' anisotropy is robust~\cite{copi:krauss,ck2,pap1,kuhlen}.

A sufficiently high density, high speed tidal stream would produce a step in the differential event rate~\cite{gg}, the position and height of which vary annually~\cite{savage}. In a directional experiment a stream would produce recoils which are strongly peaked in the opposite direction~\cite{ag2}, allowing the stream properties to be measured~\cite{OHare:2014nxd}. A dark disc with a sufficiently high density would increase the differential event rate at low energies and also change the phase and amplitude of the annual modulation signal~\cite{Bruch:2008rx}.

\section{How to handle the uncertainties?}
\label{strategies}
In light of the uncertainties in the local WIMP density and velocity distribution discussed in Sec.~\ref{astro1} and \ref{astro2} above, it is crucial to develop methods for analysing and comparing direct detection data that are independent of these uncertainties. Conventionally direct detection experimental results are presented in terms of constraints on the WIMP mass $m_{\chi}$ and cross-sections, $\sigma^{\rm SI}$ and $\sigma^{\rm SD}$, however this requires assumptions to be made about the local density and velocity distribution.  The energy dependence of the differential event rate depends on both the WIMP mass and velocity distribution, so that it is impossible to measure the WIMP mass with a single experiment without making assumptions about the velocity distribution~\cite{dreesshan,Peter:2009ak}. In Sec.~\ref{indep} we discuss astrophysics independent methods, in Sec.~\ref{param} we explore parameterizing the WIMP speed distribution and finally, in Sec.~\ref{mod}, we discuss using astrophysical data to constrain mass models of the MW.

\subsection{Astrophysics independent methods}
\label{indep}
In principle, with data from two experiments using different targets, the WIMP mass can be be determined, without any assumptions about $f(v)$ by taking moments of the energy spectra~\cite{dreesshan,Peter:2009ak}. However the WIMP mass is systematically underestimated by this method if it is comparable with or heavier than the mass of the target nuclei.

Fox et al.~\cite{Fox:2010bz} showed that an astrophysics independent comparison of different experiments can be made using the rescaled velocity integral, $\tilde{g}(v_{\rm min})= (\rho_{0} \sigma_{\rm p}^{\rm SI}/ m_{\chi}) g(v_{\rm min})$, where $g(v_{\rm min})$ is defined in Eq.~(\ref{gvmin}). Different experiments probe different sections of the speed distribution, and it is possible to assess whether they overlap and whether or not the measured rates are consistent. Because the relationship between recoil energy $E$ and $v_{\rm min}$ depends on the (unknown) WIMP mass, the comparison has to be done for a selection of fixed values of the WIMP mass.
Alternatively the comparison can be done in terms of the nuclear recoil momentum which, unlike $v_{\rm min}$, is independent of $m_{\chi}$~\cite{Anderson:2015xaa}. 

The original method has been generalized to include annual modulation~\cite{Frandsen:2011gi,Gondolo:2012rs,HerreroGarcia:2012fu,DelNobile:2013cta}, experiments containing multiple isotopes~\cite{Gondolo:2012rs}, energy-dependent experimental efficiency~\cite{DelNobile:2013cta} as well as `non-standard' particle interactions~\cite{Bozorgnia:2013hsa,Miao:2013sqa,Shan:2011ka,DelNobile:2013cva,Cherry:2014wia,Kahlhoefer:2016eds}. This method is an extremely powerful tool for examining whether signals and limits from different experiments are consistent with each other. In particular it has been used to show that the excesses over backgrounds seen by CRESST~\cite{Angloher:2011uu}, CoGeNT~\cite{Aalseth:2012if} and CDMS Si~\cite{Agnese:2013rvf}, and the annual modulations seen by DAMA/LIBRA~\cite{Bernabei:2013xsa} and CoGeNT~\cite{Aalseth:2014eft} are not compatible with each other and also the exclusion limits from LUX~\cite{Akerib:2016vxi}, PandaX~\cite{Tan:2016zwf} and CDMSlite~\cite{Agnese:2015nto} in the absence of fine-tuned non-standard interactions (see e.g. Ref.~\cite{Fox:2013pia}).

This method has been developed so that it can be applied to non-binned data~\cite{Fox:2014kua,Feldstein:2014ufa} and the degree of agreement between different experiments quantified~\cite{Feldstein:2014gza,Feldstein:2014ufa,Gelmini:2015voa,Gelmini:2016pei}. This involves writing the rescaled velocity integral as a series of monotonically decreasing steps (which is physically equivalent to the velocity distribution being composed of a series of streams with different densities and velocities) and finding the optimum step properties~\cite{Feldstein:2014gza,Fox:2014kua,Feldstein:2014ufa,Kahlhoefer:2016eds,Ibarra:2017mzt}.  It is also possible to make astrophysics independent comparisons of direct detection experiments and neutrino telescopes~\cite{Ferrer:2015bta,Ibarra:2017mzt}.

\subsection{Parameterizing the speed distribution}
\label{param}

If the underlying shape of the speed distribution is known, it is possible to make unbiased reconstructions of the WIMP particle physics parameters by marginalising over the parameters of the speed distribution (for instance the peak speed)~\cite{Peter:2009ak,Pato:2010zk}.
However Peter~\cite{Peter:2011eu} showed that erroneous assumptions about the shape of the local speed distribution lead to biased determinations of the WIMP particle physics parameters. She suggested the alternative approach of using an agnostic empirical parameterization of the speed distribution, and jointly constraining this, along with the WIMP particle parameters, using data from multiple experiments. A simple parameterization, as a series of bins with constant values, performed better than assuming an incorrect functional form, however the reconstructed parameter values were still biased~\cite{Peter:2011eu}. Kavanagh and Green~\cite{Kavanagh:2012nr} showed that this was because with fixed speed bins, the width of the bins in energy space can be reduced, so as to give a better fit to the measured energy spectrum, by artificially reducing the WIMP mass. They proposed parameterising the momentum distribution instead to avoid this problem. 
They subsequently found~\cite{Kavanagh:2013wba} that parameterizing the natural logarithm of the speed distribution in terms of Legendre polynomials allows an accurate, unbiased reconstruction of the WIMP mass to be made, even for speed distributions containing a high density stream or dark disc. It is also possible to reconstruct the speed distribution itself, however the uncertainties are quite large~\cite{Kavanagh:2013wba}.

It is not possible to make unbiased measurements of the cross-sections using direct detection experiments alone, however, as we do not know what fraction of the WIMP have speeds which are too low to cause recoils with energies above the experimental threshold energy. Ref.~\cite{Kavanagh:2014rya} showed that this problem can be avoided by combining data from direct detection experiments with neutrino telescope data. This is because it is the low speed WIMPs, that direct detection experiments are not sensitive to, that are captured in the Sun and then annihilate. So by combining these data sets the entire speed distribution is probed. Unbiased measurements of both the WIMP mass and cross sections can then be made using either the binned or polynomial speed parameterisations. The polynomial parameterisation allows for a wider range of possible speed distributions, and hence leads to larger uncertainties in the particle physics parameters.  These methods have recently been extended to parameterizing the full 3d velocity distribution as probed by directional experiments~\cite{Kavanagh:2015aqa}, see also Refs.~\cite{Alves:2012ay,Lee:2014cpa}.

See Ref.~\cite{Peter:2013aha} for a detailed study of how the WIMP astrophysics and particle physics can be constrained using multiple direct detection experiments.

\subsection{Modelling the Milky Way}
\label{mod}

In early work in this area Strigari and Trotta~\cite{Strigari:2009zb} constructed a mass model of the MW and used stellar kinematics and (simulated) direct detection data to simultaneously fit the WIMP particle physics parameters and the free parameters of the speed distribution, which they assumed to be an isotropic Maxwellian. Catena and Ullio~\cite{Catena:2011kv} built mass models of the MW, including its luminous components and a range of possibilities for the density profile of the DM halo. They used a large ensemble of astronomical data sets to constrain the parameters of the mass model, and then used the Eddington formalism to self-consistently calculate the speed distribution (see also Ref.~\cite{Pato:2012fw}). Ref.~\cite{Fornasa:2013iaa} extended this approach to include anisotropy of the velocity distribution. The resulting mean velocity distribution has a high-speed tail that is enhanced relative to the isotropic case and is similar to the empirical form, Eq.~(\ref{maofv}), found by Mao et al. to give a good fit to N-body simulations~\cite{Mao:2012hf}. 
This approach can be extended to include information about the shape of the density profile from gamma-rays produced by WIMP annihilation in the MW halo~\cite{Cerdeno:2016znc}.

\section{Summary}

The differential event rate in WIMP direct detection experiments depends on the local dark matter density and velocity distribution. Experimental data analyses usually assume a simple Standard Halo Model, an isothermal sphere with isotropic Maxwellian speed distribution. However this model may not be an accurate description of the actual dark matter distribution in the Milky Way. We first reviewed observational determinations of the relevant quantities: the local density, circular speed and escape speed. Typically the statistical errors can be fairly small, $\sim 10 \%$, however the systematic errors may be substantially larger. We then turned our attention to numerical simulations. There has recently been significant progress in carrying out hydrodynamical simulations including baryons and it appears that the speed distribution is closer to a Maxwellian distribution than previously thought. However there may still be non-negligible deviations and significant variation from halo to halo

We then discussed the consequences of these astrophysical uncertainties. The energy spectrum depends on an integral over the WIMP speed distribution, therefore exclusion limits are usually relatively insensitive to its exact shape. However the annual modulation and detailed direction dependence are far more dependent on the detailed form of the velocity distribution. Finally we concluded by discussing methods for handling astrophysical uncertainties. Astrophysics independent methods allow model independent comparisons of data sets from different experiments.  Alternatively with an appropriate parameterization of the speed distribution it will be possible, using data from multiple experiments, to make an unbiased measurement of the WIMP mass without knowledge of, or assumptions about, the speed distribution. Finally another approach is to build a mass model of the Milky Way, constrain its parameters using observational data and calculate the speed distribution self-consistently. In the event of convincing signals being observed in direct detection experiments all three of these approaches will likely be useful in establishing their consistency and making reliable measurements of the WIMP particle physics parameters.

\section*{Acknowledgments}

The author acknowledges support from STFC grant ST/L000393/1 and the Leverhulme Trust and is grateful to David Cerdeno, Mattia Fornasa, Bradley Kavanagh, Ben Morgan, Ciaran O'Hare and Miguel Peiro for collaborations on some of the work discussed.

\end{document}